\documentclass[twocolumn,
preprintnumbers,prl,nofootinbib,longbibliography]{revtex4}

\usepackage{xcolor}
\usepackage{amsmath,amssymb}
\usepackage{hyperref}
\usepackage{graphicx}
\usepackage{times}
\usepackage{dcolumn}

\begin{document}

\title{Large $N$ twisted partition functions in 3d-3d correspondence 
	and Holography}

\author{Dongmin Gang}
\affiliation{Center for Theoretical Physics, Seoul National University, Seoul 08826, Korea}
\author{Nakwoo Kim}
\affiliation{Department of Physics and Research Institute of Basic Science,
	Kyung Hee University, Seoul 02447, Korea\\
School of Physics, Korea Institute for Advanced Study, Seoul 02445, Korea}

\date{\today}

\begin{abstract}
We study the large $N$ limit of twisted partition functions on $\mathcal{M}_{g,p}$, the $S^1$ bundle of degree $p$ over a  Riemann surface of genus $g$, for 3D $\mathcal{N}=2$ superconformal field theories arising as low-energy limit of wrapped $N$ M5-branes on
hyperbolic 3-manifold $M$. We study contributions from two  Bethe vacua  which correspond to two canonical irreducible $SL(N, \mathbb{C})$ flat connections on $M$ via 3D-3D correspondence. Using mathematical results on  perturbtaive Chern-Simons invariants around the flat connections,  we find universal expressions for  the large $N$ twisted partition functions contributed from the two Bethe vacua in term of the hyperbolic volume of $M$. The two large $N$ partition functions perfectly match the on-shell actions for two Bolt-type  solutions  in the holographic dual $AdS_4$ gravity  respectively.
\end{abstract}

\maketitle

\section{Introduction and Summary}
\label{sec:intro}
As a consistent theory of quantum gravity, string/M-theory is expected to provide microscopic understandings of  quantum aspects of black holes (BHs). In a celebrated work \cite{Strominger:1996sh}, this hope was realized for extremal black holes in Minkowski spacetime, by reproducing the Bekenstein-Hawking entropy through counting D-brane bound states. Recently, this success was extended to black holes in asymptotically anti-de-Sitter spacetime. In particular, the entropy of magnetically charged supersymmetric black holes in $AdS_4$ supergravity can be explained using the holographic principle \cite{Benini:2015eyy,Benini:2016rke,Hosseini:2016tor,Hosseini:2016ume,Cabo-Bizet:2017jsl,Azzurli:2017kxo,Hosseini:2017fjo,Benini:2017oxt,Toldo:2017qsh,Bobev:2017uzs}. The AdS/CFT correspondence \cite{Maldacena:1997re} says quantum gravity in asymptotically $AdS_{4}$ spacetime should be dual to a conformal field theory (CFT) on the $3$-dimensional (3D) boundary. The field theories of interest have $\mathcal{N}\geq 2$ supersymmetry, and the black hole entropy on the gravity side turns out to be related to the so-called {\em topologically twisted indices} \cite{Gukov:2015sna,Benini:2015noa,Benini:2016hjo,Closset:2016arn} in the dual field theory. They are  the partition functions (ptns) on supersymmetric curved backgrounds $\mathcal{M}_{g,p=0}:=\Sigma_g \times S^1$, with an appropriately chosen background magnetic flux coupled to R-symmetry current. The magnetic flux is turned on along the Riemann surface $\Sigma_g$ of genus $g$.
 The twisted indices can be then computed using the supersymmetric localization technique. Recently, the realm of localizable 3D manifolds has been further extended in \cite{Closset:2017zgf}, and we now have formulae for twisted partition functions $Z_{g,p}$, i.e. the ptn on degree-$p$ $S^1$-bundles $\mathcal{M}_{g,p}$ (see \eqref{M-g,p}) over $\Sigma_g$.
 
 In this letter, we study  holographic duality for a large class of  3D $\mathcal{N}=2$ SCFTs $T_{N}[M]$, defined in \eqref{T-N-M} (see also \eqref{T-N-M-2}),  arising from wrapped M5-branes on closed hyperbolic 3-manifolds $M$. The 3D theory is characterized
 by $N$, number of M5-branes,  and the choice of a 3-manifold $M$. For each  $M$, there is an associated AdS$_4$/CFT$_3$ correspondence. The holographic dual of the wrapped M5-brane 3D SCFT was studied in  \cite{pernici1985spontaneous,Gauntlett:2000ng}.  Since there are infinitely many such 3-manifolds \cite{thurston1979geometry}, the wrapped M5-branes system provides a huge set of AdS$_4$/CFT$_3$ examples.
We probe the holography using the twisted ptns. For $p=0$, the ptn becomes a twisted index which counts ground states of M5-branes on $ \Sigma_g \times M$. The counting is holographically dual to the microstates counting for a supersymmetric BH solution interpolating the asymptotic $AdS_4$ and its near-horizon limit $AdS_2\times \Sigma_g$ \cite{Bobev:2017uzs}.

The 3D-3D correspondence \cite{Dimofte:2010tz,Terashima:2011qi,Dimofte:2011ju} provides a novel way of analyzing the 3D $\mathcal{N}=2$ SCFTs. Schematically,  the correspondence says
 \begin{align}
 \begin{split}
&(\textrm{supersymmetric ptns of 3D $T_{N}[M]$ theory})
\\
&=(\textrm{$SL(N,\mathbb{C})$ Chern-Simons theory invariants on  $M$})\;.
\end{split}
 \end{align}
Refer to \cite{Yagi:2013fda,Lee:2013ida,Cordova:2013cea,Beem:2012mb,Dimofte:2014zga,Gukov:2016gkn,Gukov:2017kmk,Mikhaylov:2017ngi} for more details on 3D-3D  dictionary. Via the correspondence, some supersymmetric quantities of 3D $T_{N}[M]$ theory can be evaluated without relying on a field theoretic description of the 3D SCFT. For example, as summarized in Table~\ref{3d/3d correspondence for twisted ptns}, the twisted partition function $Z_{g,p}$  can be written in terms of  basic perturbative invariants of the complex Chern-Simons (CS) theory.  These  invariants are mathematically well-defined and have been extensively studied in math literature. 

Combining  the 3D-3D dictionary with mathematical results, we obtain the large $N$ behavior of the twisted partition functions contributed from  two distinguished Bethe vacua in the $T_{N}[M]$ theory. 
The Bethe vacua of our interest correspond to two irreducible $SL(N,\mathbb{C})$ flat connections on $M$ in 3D-3D dual complex CS theory. These two flat connections  also have a natural interpretation in terms of hyperbolic geometry, see eq~\eqref{two irreducible flat connections}. They also give global minimum and maximum of the absolute value of the fibering operator $\mathcal{F}$ appearing in the ptn computation \eqref{Twisted ptns from sum}, see \eqref{Bounds on F}. We confirm that the large $N$ twisted ptns from the two Bethe-vacua nicely match the on-shell action for the two Bolt-type  solutions \cite{Toldo:2017qsh} in the gravity dual respectively. The comparison is summarized in Table~\ref{large N computations}, which is the main point of this letter.

\section{3D  $T_{N}[M]$ theory }
A large class of 3D $\mathcal{N}=2$ SCFTs can be engineered through a twisted compactification of  6-dimensional SCFTs. They are labelled by the internal 3-manifolds $M$:
\begin{align}
\begin{split}
&T_{N}[M] := (\textrm{Effective 3D $\mathcal{N}=2$ SCFT obtained from }
\\
&\;\; \textrm{a twisted compactification of 6D $A_{N-1}$ $(2,0)$ theory}
\\
&\;\; \textrm{on  a 3-manifold $M$)}\;. \label{T-N-M}
\end{split}
\end{align} 
 For simplicity, we assume  $M$ is a closed  (compact) hyperbolic 3-manifold without boundary. To preserve supersymmetry, we perform a partial topological twisting along the internal 3-manifold using  $SO(3)$ vector subgroup of $SO(5)$ R-symmetry of the 6D theory.  The twisting preserves a quarter of supersymmetries and the resulting 3D theory becomes a 3D $\mathcal{N}=2$ SCFT with 4 supercharges.
 The 6D theory describes the low-energy effective world-volume theory of $N$ coincident M5-branes in M-theory, and the 3D theory can be considered as an effective world-volume theory of $N$ coincident M5-branes wrapped on the compact 3-manifold $M$, i.e.
  \begin{align}
  \begin{split}
  & \textrm{$N$ coincident M5-branes on $\mathbb{R}^{3}\times M$}
  \\
  & \textrm{in M-theory on $\mathbb{R}^{3} \times (T^*M) \times \mathbb{R}^2$}
  \\
  & \xrightarrow{\textrm{  \;\;\; IR world-volume theory of M5-branes\;\;\;  } } \textrm{$T_N[M]$ on $\mathbb{R}^{3}$}\;. \label{T-N-M-2}
  \end{split}
  \end{align} 
  Here  $T^*M$ is the cotangent bundle of $M$, which is a local Calabi-Yau.
  
  Let us comment on a subtle point in the setup. As emphasized in \cite{Gang:2018wek}, in taking the twisted compactification   we need to choose a connected subset of the vacuum moduli-space of the theory defined on $\mathbb{R}^3$, in order to have a genuine 3D SCFT.  For a hyperbolic $M$, there is a natural choice (which is actually a single point)  which is expected to become a discrete set of vacua when the theory is put on $\mathbb{R}^2\times S^1$. This discrete set of vacua corresponds to a subset of  irreducible $SL(N,\mathbb{C})$ flat connections on $M$. A field theoretic construction of the effective 3D gauge theory  is proposed in  \cite{Gang:2018wek},  extending  the beautiful construction in \cite{Dimofte:2011ju,Dimofte:2013iv} for cusped 3-manifolds with at least  one torus boundary component, by incorporating gauge theoretical operations corresponding to Dehn filling (removing torus boundaries) operations  on 3-manifold.

\section{$T_{N}[M]$ on $\mathcal{M}_{g,p}$ for even $p$}

We now turn to the case where $T_{N}[M]$ is put on a large class of nontrivial backgrounds $\mathcal{M}_{g,p}$ \cite{Closset:2017zgf}:
\begin{align}
\begin{split}
&\mathcal{M}_{g,p}:= (\textrm{$S^1$-bundle of degree $p$}
\\
& \qquad \qquad \textrm{over a Riemmann surface $\Sigma_g$ of genus $g$})\;,
\\
& \textrm{i.e. } S^1 \xrightarrow{\;  p \;} \mathcal{M}_{g,p} \rightarrow \Sigma_g \;. \label{M-g,p}
\end{split}
\end{align}  
%
The metric can be written as 
\begin{align}
ds^2 = \beta^2 (d \psi - p a(z,\bar{z}))^2 + 2 g_{z \bar{z}} dz d \bar{z}\;,
\end{align}
where $z,\bar{z}$ are local coordinates on the Riemann surface and $\psi \sim \psi+2\pi$ parameterizes the $S^1$-fiber of length $\beta$. $a$ is a 1-form on $\Sigma_g$ whose curvature $F_a :=da$ is normalized as
\begin{align}
\frac{1}{2\pi} \int_{\Sigma_g} d a =1\;.
\end{align} 
To preserve some supersymmetry, we turn on the following background gauge field coupled to $U(1)$ R-symmetry. 
\begin{align}
A^{R}=  \beta \nu_R (d \psi - p a) + n_R (\pi^*  a)\;, \label{background U(1)-R}
\end{align}
with proper quantization conditions for $(\nu_R, n_R)$ \cite{Toldo:2017qsh}. Here $\pi^* a $ is a 1-form on $\mathcal{M}_{g,p}$ given as the pull-back of $a$ using the projection map $\pi : \mathcal{M}_{g,p} \rightarrow \Sigma_g $. 

For later comparison with the bolt-type solutions in the supergravity, we follow \cite{Toldo:2017qsh} and choose
\begin{align}
\nu_R = \frac{1}2\;, \quad n_R = \frac{p}2 + g-1\;, \quad p \in 2 \mathbb{Z}\;. \label{background U(1)-R-2}
\end{align}  

Throughout the letter,  we restrict our attention to the choice in \eqref{background U(1)-R-2} and some formulae below may not work for other cases.  For example, the large $N$ computation in Table \ref{large N computations} give incorrect answer for the usual round $S^3$ case which is $\mathcal{M}_{g=0,p=1}$.

For small $N$ the effective 3d theory $T_{N}[M]$ might witness emergent symmetries in addition to R-symmetry, as pointed out in \cite{Gang:2017lsr,Gang:2018wek}.  
When $N$ is large enough, on the other hand,  there is no accidental symmetry  and the $U(1)$ R-symmetry in the IR should be simply inherited from the compact $SO(2)$ subgroup of $SO(5)$ R-symmetry in the   6D  theory. It implies that the $U(1)$ R-charge, $R$, should be properly quantized
\begin{align}
R (\mathcal{O}) \in \mathbb{Z}\;, \; \textrm{for any state  $\mathcal{O}$ of $T_{N}[M]$ on $\Sigma_g$}\;. \label{quantization of R}
\end{align}
The Dirac quantization conditions for the $U(1)$ R-symmetry flux on $\Sigma_g$ are
\begin{align}
\begin{split}
&R(\mathcal{O})\times n_R = R(\mathcal{O})\times \big{(} \frac{p}2 + g-1\big{)} \in \mathbb{Z} \;, 
\\
&\; \textrm{for any state  $\mathcal{O}$ of $T_{N}[M]$ on $\Sigma_g $}\;.
\end{split}
\end{align} 
From \eqref{quantization of R}, we see that the Dirac quantizations are always satisfied for even $p$. 
In summary, for large enough $N$ we can put the 3D $T_{N}[M]$ theory on any $\mathcal{M}_{g \in \mathbb{Z}_{\geq 0},p \in 2\mathbb{Z}}$ with supersymmetry preserving background gauge field, given in \eqref{background U(1)-R} and \eqref{background U(1)-R-2}, coupled to the R-symmetry in the IR. 

\section{Holographic dual of   $T_{N}[M]$  }
The gravity dual description is given by the uplift of a certain magnetically charged $AdS_4$ solution in the maximally supersymmetric $D=7$ gauged supergravity 
\cite{pernici1985spontaneous,Gauntlett:2000ng}. Schematically, the $D=11$ solution is a product of $AdS_4$, hyperbolic 3-manifold $M$, and a squashed 4-sphere $\tilde{S}^4$. Consistency of the truncation from $D=11$ down to minimal ${\cal N}=2, D=4$ gauged supergravity is established in \cite{Donos:2010ax} and it is guaranteed that we may replace the $AdS_4$ part with any nontrivial $D=4$ solution and we still have an exact $D=11$ solution. 

The computation of holographic free energy can be also first done in $D=4$ setup, and substitute the Newton constant with \cite{Gang:2014ema}
\begin{align}
G_4 = \frac{3\pi^2}{2N^3 \textrm{vol}(M)}\;.  \label{4d G}
\end{align}
Here, the hyperbolic volume is defined as
\begin{align}
\begin{split}
&\textrm{vol}(M) = (\textrm{hyperbolic volume of $M$})
\\
& :=(\textrm{volume measured in the unique  hyperbolic metric}) \;.\nonumber
\end{split} 
\end{align}
The hyperbolic metric is normalized as $R_{\mu\nu} = - 2 g_{\mu\nu}$. The Mostow's rigidity theorem \cite{mostow1968quasi} guarantees the uniqueness of  the hyperbolic metric and thus the volume is actually a topological invariant. 

As gravity duals of the boundary theory put on ${\cal M}_{g,p}$, we utilize the supersymmetric AdS-Taub-NUT and bolt solutions constructed in \cite{Martelli:2012sz}. Since these solutions have non-vanishing Maxwell field, which in $D=11$ uplift appears as a twisting of the R-symmetry angle in $\tilde{S}^4$, one might worry about a conflict with the quantization condition $g,p$. But it turns out, since the R-symmetry angle is part of $\tilde{S}^4$ and we have a standard periodicity of $2\pi$, the regularity condition for $D=4$ NUT/Bolt is enough. 
This is in line with the field theory side discussion, in particular \eqref{quantization of R}. A comment is in order here, in comparison with the uplifts involving Sasaki-Einstein 7-manifolds. In that case, the periodicity of the R-symmetry angle from the regularity of Bolt solution should be compatible with the periodicity condition due to collapsing cycles in the K\"ahler-Einstein base manifold of the Sasaki-Einstein space. The readers are referred to \cite{Toldo:2017qsh} for more details, where the authors considered an explicit example of Sasaki-Einstein manifolds such as $V^{5,2}=SO(5)/SO(3)$.

\section{Twisted partition functions of $T_N[M]$ in 3D-3D correspondence }
The twisted partition function $Z_{g,p}$ on the $\mathcal{M}_{g,p}$ for general 3D $\mathcal{N}=2$ SCFTs is given as the following finite sum \cite{Gukov:2016gkn,Closset:2017zgf}
\begin{align}
Z_{g,p} =  \sum_\alpha Z^{\alpha}_{g,p} := \sum_\alpha (\mathcal{H}^\alpha)^{g-1} (\mathcal{F}^\alpha)^p\;. \label{Twisted ptns from sum}
\end{align}
Here $\alpha$ labels the so-called Bethe vacua \cite{Nekrasov:2014xaa} of the 3D theory. It is obtained by extremizing the effective 2d twisted superpotential in the compactification on $\mathbb{R}^2 \times S^1$.
The number of vacua is equal to the Witten index \cite{Kim:2010mr,Intriligator:2013lca} of the 3D SCFT. 
$\mathcal{H}$ and $\mathcal{F}$ are called {\it handle-gluing} and {\it fibering} operators respectively. The explicit forms of $\mathcal{H}$ and $\mathcal{F}$ for any given ultra-violet (UV) Lagrangian are available in \cite{Closset:2017zgf}. Let us emphasize that, the formula \eqref{Twisted ptns from sum} applied to the case of $S^3$ partition function which corresponds to $(g,p)=(0,1)$ is apparently {\it different} from the more familiar  Coulmob branch integral expression \cite{Kapustin:2009kz}. But their equivalence is illustrated for a number of examples in \cite{Closset:2017zgf}.
\begin{table}[h]
	\begin{center}
		\begin{tabular}{|c|c|}
			\hline 
			3D $T_{N}[M]$ theory & $SL(N, \mathbb{C})$ CS theory \\
			\hline \hline
			Bethe vacuum  $\alpha$ & \;\; Irreducible  flat connection $\mathcal{A}^{\alpha}$\;\; \\ 
			\hline 
			Handle gluing operator $\mathcal{H}^{\alpha}$ & $\exp (-2 S_1^{\alpha} )$  \\ 
			\hline 
			Fibering  operator $\mathcal{F}^{\alpha}$ & $\exp (i S_0^{\alpha}/(2\pi) )$   \\ 
			\hline 
		\end{tabular} \caption {3d-3d dictionaries for basic ingredients in twisted partition function computation. 
			$S^{\alpha}_{n=0,1}$ are perturbative invariants of the complex Chern-Simons theory around a flat-connection $\mathcal{A}^\alpha$, see eq~\eqref{S0-S1}.  }
		\label{3d/3d correspondence for twisted ptns}
	\end{center}
\end{table}

Now let us specialize to the $T_{N}[M]$ theories  in \eqref{T-N-M}. The twisted ptns for these theories can be analyzed using the  3D-3D dictionaries  summarized in Table~\ref{3d/3d correspondence for twisted ptns}.  Twisted ptns in 3D-3D correspondence were studied in \cite{Gukov:2016gkn,Gukov:2017kmk}.
 In the table,  the $\{S_n^{\alpha}\}_{n=0}^\infty$ represent terms in the loop expansion of the complex  CS  partition function around a flat-connection $\mathcal{A}^{\alpha}$ \cite{Gukov:2003na,Dimofte:2009yn,Gukov:2006ze,Gang:2017cwq}: 
\begin{align}
\begin{split}
&Z_{\rm CS \; pert}^{\alpha} :=\int \frac{D (\delta \mathcal{A})}{(\textrm{gauge})} e^{- \frac{1}{2\hbar} CS [\mathcal{A}^{\alpha} + \delta \mathcal{A};M]} 
\\
&\xrightarrow{\textrm{\;\; as $\hbar$ goes to  0}\;\;} \exp \left(\frac{1} \hbar S^{\alpha}_0 + S^{\alpha}_1  + \ldots + \hbar^{n-1} S^\alpha_n + \ldots \right)\;.
\end{split}
\end{align}
The  Chern-Simons functional is 
\begin{align}
CS[\mathcal{A},M]:= \int_M \textrm{Tr} (\mathcal{A}\wedge d \mathcal{A} + \frac{2}3 \mathcal{A}^3)\;. \label{S0,S1 and S2}
\end{align}
Note that the counterpart of ${\cal F}$ and ${\cal H}$ are simply {\it tree level} and {\it one-loop} contributions in perturbation theory!
More explicitly, the perturbative coefficients are given as
\begin{align}
\begin{split}
&S_0^{\alpha} = - \frac{1}2 CS [\mathcal{A}^\alpha,M] \;,
\\
&S_1^{\alpha} := \frac{1}2 \log \textrm{Tor}_{R={\rm adjoint}}[\mathcal{A}^\alpha, M]\;.
\end{split} \label{S0-S1}
\end{align}
  $\textrm{Tor}_R [\mathcal{A}^\alpha, M]$ is the Ray-Singer torsion of an associated vector bundle in a representation $R \in \textrm{Hom}[SL(N,\mathbb{C}) \rightarrow {GL}(V_R)]$  twisted by a flat connection $\mathcal{A}^{\alpha}$. Here $V_R$ is the vector space for representation $R$ and $GL(V_R)$ is the general linear group on the $V_R$. The analytic torsion is defined as follows \cite{ray1971r,Gukov:2006ze,Gukov:2011qp}
\begin{align}
\textrm{Tor}_{ R} [ \mathcal{A}^\alpha, M] := \frac{[\textrm{det}'\Delta_0 (R, \mathcal{A}^\alpha)]^{3/2}}{[\textrm{det}'\Delta_1 (R, \mathcal{A}^\alpha)]^{1/2}}\;.
\end{align}
Here $\Delta_n (R, \mathcal{A}^\alpha )$ is a Laplacian acting on $V_R$-valued
$n$-form twisted by a flat connection $\mathcal{A}^\alpha$. ${\rm det}' \Delta$ denotes the zeta function regularized determinant of the Laplacian $\Delta$. For the one-loop part, the denominator comes from gauge field fluctuations $\delta  \mathcal{A}$ while the numerator comes from the ghosts associated to a gauge fixing \cite{witten1989quantum}.

The 3D-3D dictionary in Table~\ref{3d/3d correspondence for twisted ptns} can be derived combining several  known results in literatures. The Bethe-vacua (vacua on $\mathbb{R}^2 \times S^1$)  of 3D $T_{N}[M]$ theory  are in one-to-one correspondence to a  subset  of irreducible flat-connections on $M$ \cite{Dimofte:2010tz,Gang:2018wek}. According to a dictionary of 3D-3D relation, the asymptotic expansion $Z^{\alpha}_{\rm CS \; pert}$ in  \eqref{S0,S1 and S2} is equal to  the perturbative expansion of holomorphic block $B^{\alpha} (q)$ \cite{Dimofte:2010tz,Beem:2012mb} associated to the Bethe-vacuum $\alpha$ in the limit $q \rightarrow 1$,
\begin{align}
\begin{split}
&Z^{\alpha}_{\rm CS \; pert} (\hbar) \simeq B^{\alpha} (q:=e^{\hbar}),
\\
&\textrm{as an asymptotic expansion in $\hbar \rightarrow 0$}.
\end{split}
\end{align} 
For general 3D $\mathcal{N}=2$ theory, the asymptotic expansion coefficients $S_0$ and $S_1$ of holomorphic block  are related to the operators $\mathcal{F}$ and $\mathcal{H}$ as given in  Table~\ref{3d/3d correspondence for twisted ptns} \cite{Closset:2018ghr,to-appear}.

\section{Large N twisted partition functions around two Bethe-vacua and its holographic dual}
For every hyperbolic 3-manifold $M$, there are two characteristic irreducible $SL(N, \mathbb{C})$ flat connections $\mathcal{A}^{\rm geom}_N$ and $\mathcal{A}^{\overline{\rm geom}}_N$ which can be constructed from the  hyperbolic structure on $M$,%
\begin{align}
\mathcal{A}^{\rm geom}_N := \rho_N \cdot (\omega + i e)\;, \quad \mathcal{A}^{\overline{\rm geom}}_N := \rho_N \cdot (\omega - i e)\;.  \label{two irreducible flat connections}
\end{align}
Here $\omega$ and $e$ are respectively the spin-connection and  vielbein of the unique hyperbolic metric on $M$.  They are both locally $so(3)$-valued 1-forms and the complex combinations $\omega \pm i e$ form an $SL(2,\mathbb{C})$ flat-connections on $M$. $\rho_N$ is the $N$-dimensional irreducible representation of $sl(2,\mathbb{C}) = su(2)_{\mathbb{C}}$, and obviously $\rho_N \cdot (\omega \pm  i e)$ become also irreducible $SL(N,\mathbb{C})$ flat connections. A crucial property of these two flat-connections is that they take the minimum (maximum) value of $\textrm{Im}[S_0]$ among all  $SL(N,\mathbb{C})$ flat connections. Namely,
\begin{align}
\textrm{Im}[S_0^{\overline{\rm geom}}]<\textrm{Im}[S_0^\alpha]<\textrm{Im}[S_0^{\rm geom}] \;, \label{Bounds on Im[S0]}
\end{align}
for any flat connection $\mathcal{A}^\alpha$ which is neither $\mathcal{A}^{\overline{\rm geom}}$ nor $\mathcal{A}^{\rm geom}$. Combined with the 3D-3D dictionary in Table~\ref{3d/3d correspondence for twisted ptns}, it is implied
\begin{align}
|\mathcal{F}^{\rm geom}|<|\mathcal{F}^{\alpha}|<|\mathcal{F}^{\overline{\rm geom}}| \;, \label{Bounds on F}
\end{align}
for any Bethe-vacuum  $\alpha$ which is neither $(\overline{\rm geom})$ nor $({\rm geom})$.

Classical actions $S_0^{\alpha}$ for the two connections above can be computed  as follows
\begin{align}
\begin{split}
&\textrm{Im}[S_0^{\rm geom}] = -\frac{1}2 \textrm{Im} [CS (\mathcal{A}_{N}^{\rm geom})] = -\frac{1}2 \textrm{Im} [CS (\rho_N \cdot (\omega + i e))] 
\\
& =-\frac{1}2 \frac{\textrm{Tr}[\rho_N \cdot (T^a) \rho_N \cdot (T^b)]}{\textrm{Tr}[T^a T^b]} \textrm{Im} [CS (\omega + ie)]
\\
& = -\frac{1}2 \frac{N^3 -N}{6} \textrm{Im} [CS  (\omega + i e)] =\frac{N^3-N}{6} \textrm{vol}(M)\;, \; \;\textrm{and}
\\
&\textrm{Im}[S_0^{\overline{\rm geom}}] = - \textrm{Im}[S_0^{\rm geom}]=-\frac{N^3-N}{6} \textrm{vol}(M)\;. \label{large N classical}
\end{split}
\end{align}
In the second line, $T^{a} \;(a=1,2,3)$ are Pauli  matrices and $\rho_N \cdot (T^a)$ are  generators in the $N$-dimensional irreducible representation. From a simple group theoretical fact
\begin{align}
\frac{\textrm{Tr}[\rho_N \cdot (T^a) \rho_N \cdot (T^b)]}{\textrm{Tr}[T^a T^b]}  = \frac{N^3 -N}6 \;,
\end{align}
the expected $N^3$-scaling of $T_{N}[M]$ theory follows. In the third line of \eqref{large N classical}, we use the fact that the imaginary part of Chern-Simons  functional of  $\mathcal{A} = \omega + i e$ is equal to the Einstein-Hilbert action with unit negative cosmology constant up to an overall numerical factor \cite{witten19882+}. The action for the unique hyperboilc metric is twice of the hyperbolic volume of 3-manifold with a minus sign.

The large $N$ asymptotic behavior of the 1-loop coefficients, $S_1^{{\rm geom}}$ and $S_1^{\overline{\rm geom}}$, can be analyzed using a following mathematical theorem \cite{muller2012asymptotics},
\begin{align}
\begin{split}
&\log |\textrm{Tor}_{R=\rho_{2m+1}} [\mathcal{A}^{\rm geom}_{N=2},M]|
\\
&\xrightarrow{\textrm{ as $m$ goes to $\infty$ } }-\frac{1}\pi m^2 \textrm{vol}(M) +o (m)\;. \label{A math thm}
\end{split}
\end{align}
Here $\rho_{2m+1}$ is the $(2m+1)$-dimensional irreducible representation of $sl(2,\mathbb{C}) = su(2)_{\mathbb{C}}$. Combining the theorem with the following branching rule,
\begin{align}
\begin{split}
&\big{(}\textrm{adjoint of $sl(N, \mathbb{C})$} \big{)} = \bigoplus_{m=1}^{N-1} \rho_{2m+1}  \textrm{ of $sl(2,\mathbb{C})$}\;,
\\
& \textrm{when the  $sl(2,\mathbb{C})$ is embedded into $sl(N, \mathbb{C})$ via $\rho_N$.}
\end{split}
\end{align}
we have following large $N$ behavior of the 1-loop coefficients
\begin{align}
\begin{split}
&\textrm{Re}[S_1^{\rm geom}] =\frac{1}2 \log |\textrm{Tor}_{R={\rm adjoint}}[\mathcal{A}_N^{\rm geom}, M] |
\\
&= \frac{1}2 \sum_{m=1}^{N-1}  \log |\textrm{Tor}_{R=\rho_{2m+1}} [\mathcal{A}^{\rm geom}_{N=2},M]|
\\
&=- \frac{1}{2\pi} \textrm{vol}(M)\sum_{m=1}^{N-1} m^2 + o(m) =   -\frac{N^3+o(N^2)}{6 \pi} \textrm{vol}(M)\;,\;\; 
\\
&\textrm{and}
\\
&\textrm{Re}[S_1^{\overline{\rm geom}}] = \textrm{Re}[S_1^{ \rm geom}]=-\frac{N^3+o(N^2)}{6 \pi} \textrm{vol}(M)\;. \label{large N 1-loop}
\end{split}
\end{align}
Combining the 3D-3D dictionaries in Table~\ref{3d/3d correspondence for twisted ptns} with the large $N$ analysis in \eqref{large N classical} and \eqref{large N 1-loop},  we finally obtain following universal large $N$ behavior of the twisted ptns \eqref{Twisted ptns from sum}  
\begin{align}
\begin{split}
&F^{\rm geom}_{g,p} :=-\log |Z^{\rm geom}_{g,p}(T_N[M])|
\\
&= \frac{ 4(1-g) N^3+p N^3 }{12\pi} \textrm{vol}(M)+  o(N^2) \;,
\\
&F^{\overline{\rm geom}}_{g,p}:=-\log |Z^{\overline{\rm geom}}_{g,p}(T_N[M])| 
\\
&=   \frac{ 4(1-g) N^3 -pN^3  }{12\pi} \textrm{vol}(M)+  o(N^2) \;.
\end{split}
\end{align}
They nicely match the on-shell actions $I^{Bolt_\pm}_{g,p}$ of $Bolt_\pm $ solution in \cite{Toldo:2017qsh}.  The large $N$ computations are summarized in Table~\ref{large N computations}.
\begin{table}[h]
	\begin{center}
		\begin{tabular}{|c|c|}
			\hline 
			M-theory on  $AdS_4 \times M \times \tilde{S}^4$ & $SL(N, \mathbb{C})$ CS theory on $M$ 
			\\
			\hline \hline
			$Bolt_+$ solution\;  &  Flat connection $ \mathcal{A}_N^{\overline{\rm geom}}$
			\\
			$I^{Bolt_+}_{g,p} = \frac{\pi(4(1-g)-p)}{8G_4} $ &   $F^{\overline{\rm geom}}_{g,p} = \frac{ 4(1-g) N^3 -pN^3  }{12\pi} \textrm{vol}(M) $ 
			\\
			\hline 
			$Bolt_-$ solution\;  &  Flat connection $ \mathcal{A}_N^{\rm geom}$
			\\
			$I^{Bolt_-}_{g,p} = \frac{\pi(4(1-g)+p)}{8G_4} $ &   $F^{\rm geom}_{g,p} = \frac{ 4(1-g) N^3 +pN^3  }{12\pi} \textrm{vol}(M) $ 
			\\
			\hline 
		\end{tabular} \caption { The M-theory is holographic dual to 3D $T_{N}[M]$ theory while the complex Chern-Simons theory is 3D-3D dual to the $T_{N}[M]$ theory. The 4d Newton constant $G_4$ is given in \eqref{4d G}.   }
		\label{large N computations}
	\end{center}
\end{table}
\section{Comparison with large $N$ $S^3_b$-ptn}
The prescription to compute the twisted partition function $Z_{g,p}$ for $T_{N}[M]$ through 3D-3D correspondence naturally shares several ingredients with the corresponding computation of a squashed 3-sphere partition function  $Z_{b} (T_N [M])$ studied in \cite{Gang:2014qla,Gang:2014ema}. The squashed 3-sphere $S^3_b$ of our interest is a supersymmetric curved background introduced in \cite{Hama:2011ea}, defined as 
\begin{align}
S^3_b = \big{\{} (z,w) \in \mathbb{C}^2 \;:\; b^2 |z|^2 +\frac{1}{b^2}|w|^2=1  \big{\}}\;.
\end{align}  
Setting $b=1$ gives the usual round 3-sphere. 
According to the 3D-3D relation \cite{Terashima:2011qi,Dimofte:2011ju}, the extreme squashing limit $b \in \mathbb{R}\rightarrow 0$ corresponds to a weakly coupled limit of the Chern-Simons theory. More concretely $Z_b$ is determined by the perturbative invariants $S_n^{\overline{\rm geom}}$ in \eqref{S0,S1 and S2} around the flat connection $\mathcal{A}_N^{\overline{\rm geom}}$ \cite{Bae:2016jpi,Mikhaylov:2017ngi,Gang:2017hbs},
\begin{align}
\begin{split}
&F_b (T_{N}[M]):=-\log |Z_{b}(T_N[M])| 
\\
&\xrightarrow{\; \textrm{as $\hbar:= 2\pi i b^2$ goes to $0$}\;}
\\
&-\textrm{Re}\bigg{[}\left(\frac{1}{\hbar} S_0^{\overline{\rm geom}} + \ldots+ \hbar^{n-1} S_n^{\overline{\rm geom}} +\ldots \right)\bigg{]} \;. 
\end{split}
\end{align}
Combined with the large $N$ behaviors of $S_n^{\overline{\rm geom}}$ for $n=1,2$  in eqs. \eqref{large N classical} and \eqref{large N 1-loop}, we see that the asymptotic expansion is compatible with the gravity dual side of free-energy $I_{b}^{\rm gravity}$ \cite{Gang:2014qla}, 
\begin{align}
\begin{split}
&I_{b}^{\rm gravity} = \frac{\pi (b+b^{-1})^2}{8 G_4} = \frac{N^3}{12\pi }(b+b^{-1})^2 \textrm{vol}(M) 
\\
& \xrightarrow{\; \textrm{as $\hbar:= 2\pi i b^2$ goes to $0$}\;} 
\\
&\frac{i N^3 \textrm{vol}(M)}{6\hbar}  + \frac{N^3 \textrm{vol}(M)}{6\pi} - \frac{i N^3 \textrm{vol}(M)}{24 \pi^2} \hbar \;, \label{Ib-gravity}
\end{split}
\end{align}
up to $o(\hbar^0)$. Motivated from the comparison, it was further conjectured that \cite{Gang:2014qla}
\begin{align}
\lim_{N \rightarrow \infty} \frac{1}{N^3} S_2^{\overline{\rm geom}} =  \frac{i\,   \textrm{vol}(M)}{24 \pi^2}\;, \quad \lim_{N \rightarrow \infty} \frac{1}{N^3} S_{n\geq 3}^{\overline{\rm geom}} =  0 \; . 
\end{align}
This conjecture was checked numerically for a number of concrete examples. 

Now we compare the two large $N$ analysis and see
\begin{align}
\begin{split}
&\lim_{N\rightarrow \infty}\frac{1}{N^3}F_b (T_{N}[M])  
\\
&=\lim_{N\rightarrow \infty} \frac{1}{N^3} \left(- \frac{ \textrm{Im}[S^{\overline{\rm geom}}_0]}{2 \pi b^2}-\textrm{Re}[S^{\overline{\rm geom}}_1]+(2\pi b^2) \textrm{Im}[S^{\overline{\rm geom}}_2] \right)\;, 
\\
& \textrm{and}
\\
&\lim_{N\rightarrow \infty}\frac{1}{N^3}F_{g,p}^{\overline{\rm geom}}(T_N[M])
\\
& = \lim_{N\rightarrow \infty} \frac{1}{N^3} \left(2(g-1) \textrm{Re}[S_1^{\overline{\rm geom}}]) +\frac{p}{2\pi} \textrm{Im}[S_0^{\overline{\rm geom}}]\right).
\end{split}
\end{align}
From the comparison, we obtain a general relation of the following form, between the twisted and the squashed $S^3$ partition functions in the large $N$ limit
\begin{align}
\lim_{N\rightarrow \infty}\frac{F_{g,p}^{\overline{\rm geom}}}{F_{b=1}} = (1-g) -\frac{p}4\;. \label{universal large N relation}
\end{align}
which holds for every 3D $T_{N}[M]$ theory. To arrive the conclusion, we use following universal relation between perturbative invariants
\begin{align}
\lim_{N\rightarrow \infty}\frac{\pi \textrm{Re}[S_1^{\overline{\rm geom}}]}{\textrm{Im}[S_0^{\overline{\rm geom}}]} = \lim_{N\rightarrow \infty}\frac{-4\pi^2 \textrm{Im}[S_2^{\overline{\rm geom}}]}{\textrm{Im}[S_0^{\overline{\rm geom}}]}=1\;,
\end{align}
which follows from the fact   \cite{Martelli:2011fu}
\begin{align}
\begin{split}
&\lim_{N\rightarrow \infty} \frac{F_{b}}{F_{b=1}}
\\
&= \lim_{N\rightarrow \infty} \frac{-\frac{1}{2\pi}\textrm{Im}[S^{\overline{\rm geom}}_0]b^{-2}-\textrm{Re}[S^{\overline{\rm geom}}_1]+(2\pi)\textrm{Im}[S^{\overline{\rm geom}}_2]b^2}{-\frac{1}{2\pi}\textrm{Im}[S^{\overline{\rm geom}}_0]-\textrm{Re}[S^{\overline{\rm geom}}_1]+(2\pi)\textrm{Im}[S^{\overline{\rm geom}}_2]}
\\
&=\frac{1}4 (b^{-2}+ 2+b^2)\;. \nonumber
\end{split}
\end{align}
The same universal relation \eqref{universal large N relation} for $p=0$ was observed in \cite{Hosseini:2016tor,Bobev:2017uzs} for different class of 3D $\mathcal{N}=2$ SCFTs.

\section{Conclusion}
In this letter, we probe a large class of  AdS$_4$/CFT$_3$ associated to  M5-branes wrapped on 3-manifolds  by computing large $N$ twisted ptns. For M2 and D2-branes and their Chern-Simons-matter theories, the large $N$ computations have been performed already in \cite{Benini:2015eyy,Benini:2016rke,Hosseini:2016tor,Hosseini:2016ume,Cabo-Bizet:2017jsl,Azzurli:2017kxo,Hosseini:2017fjo,Benini:2017oxt,Toldo:2017qsh,Bobev:2017uzs}. A nicer feature of our analysis is that we  map the large $N$ analysis  to a mathematical problem via 3D-3D correspondence  which  can be solved from known mathematical results, such as  \eqref{Bounds on Im[S0]}, \eqref{large N classical} and \eqref{large N 1-loop}.  The results hold for any closed hyperbolic 3-manifold $M$ and one does not need to perform the large $N$ analysis for individual AdS$_4$/CFT$_3$ model associated to each $M$. We can also give a simple explanation why there is a  universal relation (see \eqref{universal large N relation}) between the twisted ptns and $S^3$-ptn in the large $N$ limit. Both types of ptns are related to the same perturbative invariants of a complex CS theory through the 3D-3D correspondence in the large $N$ limit. We hope that the improvement made in our analysis may provide a better way of understanding the sub-leading corrections to the large $N$ twisted ptns which might be related to quantum corrections to the  Bekenstein-Hawking entropy.

\section{Acknowledgments} 
We would like to thank Seok Kim, Sunjin Choi, Chiung Hwang, Jaewon Song, Masahito Yamazaki and Victor Mikhaylov for interesting discussions on related works. This work was initiated while the authors were visiting APCTP, Pohang for a workshop ``Strings, Branes and Gauge theories", 16-25 July 2018. We thank APCTP for hospitality.
The work of DG was supported by Samsung Science and
Technology Foundation under Project Number SSTBA1402-08. The research of NK was supported by  NRF grant 2015R1D1A1A09059301. 

\bibliographystyle{ytphys}
\bibliography{ref}

\end{document}